\documentclass{article}
\usepackage{amsmath}
\usepackage{amssymb}
\usepackage{mathtools}
\usepackage{bm}
\usepackage{authblk}
\usepackage{fancyhdr}
\usepackage{hyperref}
\DeclareMathOperator{\E}{\mbox{E}}
\renewcommand\vec{\bm}
\newcommand{\citep}{\cite}

\begin{document}

\title{Review of Probability Distributions for Modeling Count Data}

\author{F. William Townes}
\affil{\footnotesize Department of Computer Science, Princeton University, Princeton, NJ \\
\normalsize {ftownes@princeton.edu}
}

\lhead{Townes 2020} 
\rhead{Count and Multinomial Distributions}

\maketitle

\begin{abstract}
{
Count data take on non-negative integer values and are challenging to properly analyze using standard linear-Gaussian methods such as linear regression and principal components analysis. Generalized linear models enable direct modeling of counts in a regression context using distributions such as the Poisson and negative binomial. When counts contain only relative information, multinomial or Dirichlet-multinomial models can be more appropriate. We review some of the fundamental connections between multinomial and count models from probability theory, providing detailed proofs. These relationships are useful for methods development in applications such as topic modeling of text data and genomics.
}
\end{abstract}

\tableofcontents

\newpage
\section{Introduction}

Count data take on non-negative integer values and are challenging to properly analyze using standard linear-Gaussian methods such as linear regression and principal components analysis (PCA) \cite{hotelling_analysis_1933}. The advent of generalized linear models (GLMs) facilitated the use of the Poisson likelihood in a regression context \cite{agresti_foundations_2015}. The Poisson distribution is the simplest count model and has only a single parameter, making it unsuitable for dealing with overdispersion. This has motivated the adoption of negative binomial models, which include an additional parameter to model the dispersion separately from the mean. Negative binomial models are widely used in the analysis of counts from high-throughput sequencing experiments \cite{love_moderated_2014,hafemeister_normalization_2019}.

In parallel, the field of compositional data analysis deals with relative abundance data \cite{egozcue_isometric_2003}. The multinomial distribution is the simplest model for relative abundances when the data consist of discrete counts from each category. Like the Poisson, the multinomial cannot accommodate overdispersion; for this purpose the Dirichlet-multinomial is often used instead. Dirichlet-multinomial models are widely used in topic modeling of text \cite{blei_latent_2003} as well as in metagenomics data analysis \cite{holmes_dirichlet_2012}.

\begin{figure}[!htb]
\centering
\includegraphics[width=\linewidth]{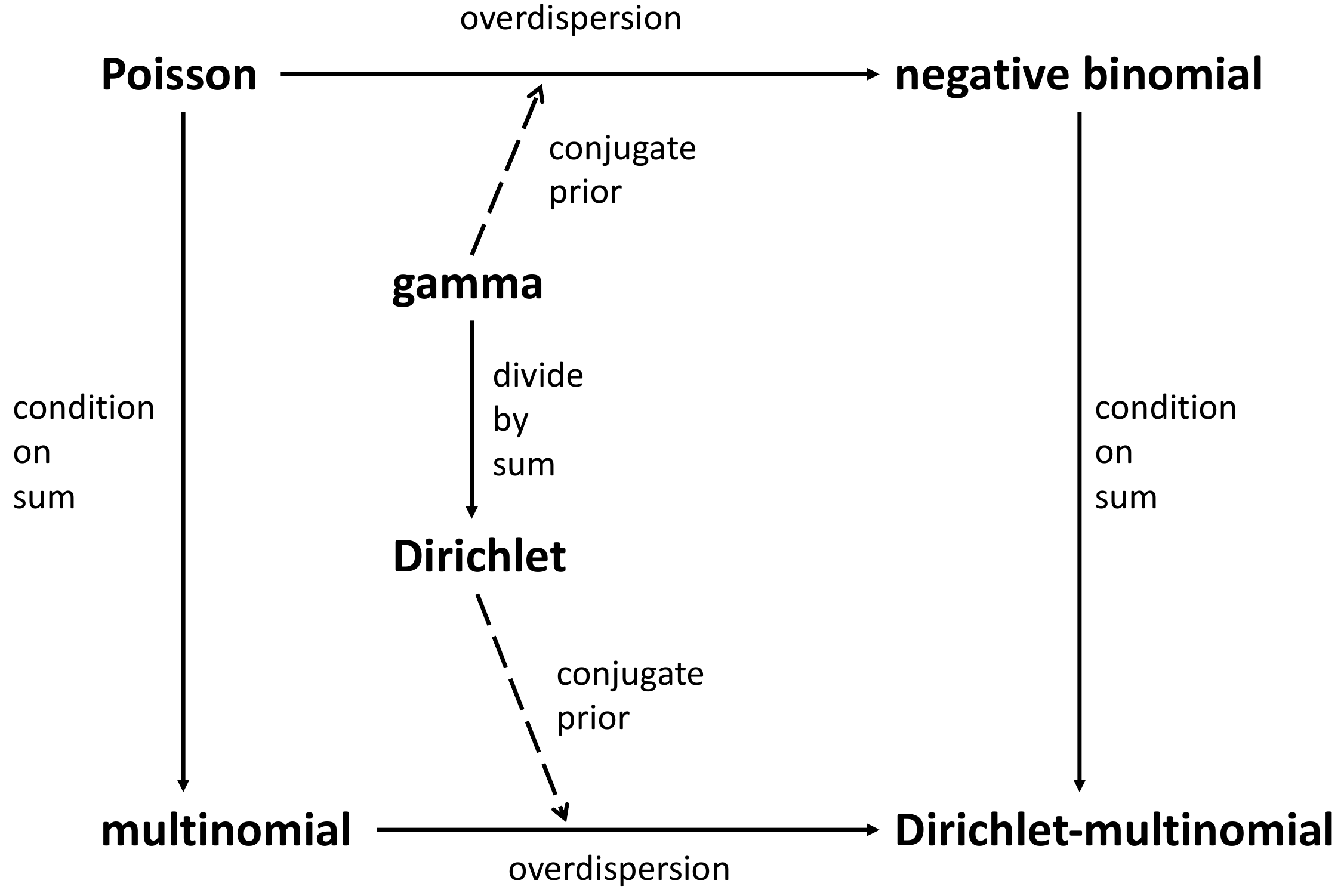}
\caption{\small Relationships between commonly used distributions for modeling of count data. All distributions are discrete except for gamma and Dirichlet.}
\label{fig:diagram}
\end{figure}

Here, we review some of the fundamental connections between multinomial and count models (Figure \ref{fig:diagram}). While none of the presented results are novel, in many cases derivations in the statistical literature have been obscure, leading to confusion among applied researchers. Here, we provide detailed proofs for all results and a discussion of the advantages and disadvantages of each distribution. A key result is the construction of the Dirichlet-multinomial from independent negative binomial distributions \cite{zhou_nonparametric_2018}. This suggests the utility of the negative binomial in modeling sequencing count data may derive from its ability to approximate the Dirichlet-multinomial, which is more realistic as a generative model for the data \citep{townes_feature_2019}.

\section{Negative binomial equivalent to Poisson-gamma}

The probability mass function (PMF) of the Poisson distribution is given by
\[p_X(x\vert \lambda)=\frac{e^{-\lambda}\lambda^x}{x!}\]
where $\E[X]=\lambda$. Suppose $\lambda$ is itself a random variable with a gamma distribution. The probability density function (PDF) is given by
\begin{align*}
f(\lambda\vert \alpha,\beta) &= \frac{\beta^\alpha}{\Gamma(\alpha)}\lambda^{\alpha-1}e^{-\beta \lambda}\\
f(\lambda\vert \alpha,\mu) &= \frac{(\alpha/\mu)^\alpha}{\Gamma(\alpha)}\lambda^{\alpha-1}e^{-\alpha\lambda/\mu}
\end{align*}
where $\E[\lambda] = \mu = \alpha/\beta$. The marginal distribution of $X$ as a Poisson-gamma random variable is formed by integrating $\lambda$ out of the joint density function.
\begin{align*}
p_X(x\vert \alpha,\mu) &= \int p(x,\lambda\vert \alpha,\mu)d\lambda = \int p_X(x\vert \lambda)f(\lambda\vert \alpha,\mu)d\lambda \\
&= \int_0^\infty \left(\frac{e^{-\lambda}\lambda^x}{x!}\right)\left(\frac{(\alpha/\mu)^\alpha}{\Gamma(\alpha)}\lambda^{\alpha-1}e^{-\alpha\lambda/\mu}\right)d\lambda \\
&= \left(\frac{(\alpha/\mu)^\alpha}{\Gamma(\alpha)x!}\right)\int_0^\infty\lambda^{(x+\alpha)-1}e^{-(1+\alpha/\mu)\lambda}d\lambda\\
&= \left(\frac{(\alpha/\mu)^\alpha}{\Gamma(\alpha)x!}\right)\frac{\Gamma(\alpha+x)}{(1+\alpha/\mu)^{(\alpha+x)}}\\
&= \frac{\Gamma(\alpha+x)}{\Gamma(\alpha)x!}\left(\frac{\mu}{\mu+\alpha}\right)^x\left(\frac{\alpha}{\mu+\alpha}\right)^\alpha
\end{align*}
This is the PMF of a negative binomial random variable with mean $\mu$ and shape parameter $\alpha$. Another convenient parametrization is obtained by setting $\theta=\frac{\mu}{\mu+\alpha}$ so that
\[p_X(x\vert \alpha,\theta) = \frac{\Gamma(\alpha+x)}{\Gamma(\alpha)x!} \theta^x(1-\theta)^\alpha\]
Under this parametrization we say that $X\sim NB(\alpha,\theta)$ and $\E[X]=\alpha\frac{\theta}{1-\theta}$. This result was first shown by \cite{greenwood_inquiry_1920}.

\section{Dirichlet as normalized sum of gamma distributions}

Let $\vec{X}=(X_1,\ldots,X_{n+1})$ be a random vector whose elements are independent gamma random variables $X_i\sim Gamma(\alpha_i,~\beta)$ such that $\E[X_i]=\alpha_i/\beta$. Let $Y=\sum_{i=1}^{n+1} X_i$ and note that $Y\sim Gamma(\alpha_0,~\beta)$ where $\alpha_0=\sum_{i=1}^{n+1} \alpha_i$. Let $Z_i = X_i/Y$ for $i=1,\ldots,n$. The joint PDF of $\vec{X}$ is given by
\[f_{\vec{X}}(\vec{x}) = \prod_{i=1}^{n+1}\frac{\beta^{\alpha_i}}{\Gamma(\alpha_i)}x_i^{\alpha_i-1}e^{-\beta x_i} = \frac{\beta^{\alpha_0}}{\prod_{i=1}^{n+1} \Gamma(\alpha_i)} e^{-\beta \sum_{i=1}^{n+1} x_i}\prod_{i=1}^{n+1} x_i^{\alpha_i-1}\]
We are interested in the joint PDF of the $Z_i$ random variables. We will first construct the joint PDF of $(\vec{Z},Y)$ as a transformation of $\vec{X}$ then integrate out $Y$. Consider the multivariate transformation $(\vec{Z},Y) = g(\vec{X})$. This is an invertible transformation and the elements of the inverse function are given by $x_i = h_i(\vec{z},y) = y z_i$ and $x_{n+1}=h_{n+1}(\vec{z},y) = y\left(1-\sum_{i=1}^n z_i\right)$. The jacobian is the determinant of the matrix of first derivatives of all elements of $h$ against all elements of $\vec{x}$. All elements of this matrix are zero except the following (where $i=1,\ldots,n$)
\begin{align*}
\frac{\partial x_i}{\partial z_i} &= y\\
\frac{\partial x_i}{\partial y} &= z_i\\
\frac{\partial x_{n+1}}{\partial z_i} &= -y\\
\frac{\partial x_{n+1}}{\partial y} &= 1-\sum_{i=1}^n z_i
\end{align*}
The $(n+1)\times(n+1)$ dimensional jacobian matrix is given by
\begin{align*}
A &= \begin{vmatrix}
y & 0 & 0 & \cdots & 0 & 0 & z_1 \\
0 & y & 0 & \cdots & 0 & 0 & z_2 \\
0 & 0 & y & \cdots & 0 & 0 & z_3 \\
\vdots & \vdots & & \ddots & \vdots & \vdots & \vdots \\
\vdots & \vdots & & & y & 0 & z_{n-1} \\
0 & 0 & 0 & \cdots & 0 & y & z_n \\
-y & -y & & \cdots & -y & -y & 1-\sum_{i=1}^n z_i
\end{vmatrix}
\end{align*}
At this point, many references simply state the value of the determinant $\vert A\vert$ (which is $y^n$) without showing how it is obtained. Here we demonstrate a step-by-step derivation for clarity. The determinant is given by the Laplace expansion along the $(n+1)$ column
\[\vert A\vert = \left(\sum_{i=1}^n (-1)^{i+(n+1)}z_i \left\vert M_{i,(n+1)}\right\vert\right)+(-1)^{(n+1)+(n+1)}\left(1-\sum_{i=1}^n z_i\right)\left\vert M_{(n+1),(n+1)}\right\vert\]
where $M_{i,j}$ denotes the $n\times n$ minor matrix formed by deleting row $i$ and column $j$ from the original matrix $A$. Note that since the bottom right minor $M_{(n+1),(n+1)}$ is a diagonal matrix with $y$ along the diagonal, its determinant is simply $y^n$. Now, consider the minor of the second to last element in the far right column we are expanding along.
\[M_{n,(n+1)} = \begin{pmatrix}
y & 0 & 0 & \cdots & 0 & 0  \\
0 & y & 0 & \cdots & 0 & 0  \\
0 & 0 & y & \cdots & 0 & 0  \\
\vdots & \vdots & & \ddots & \vdots & \vdots \\
\vdots & \vdots & & & y & 0  \\
-y & -y & & \cdots & -y & -y
\end{pmatrix}
\]
Because this is a lower triangular matrix, its determinant is also the product of its diagonal elements and $\vert M_{n,(n+1)}\vert = -y^n$. Finally, note that all the remaining minors $M_{i,(n+1)}$ can be converted to $M_{n,(n+1)}$ by column swapping. If $k$ columns are swapped this multiplies the determinant by $(-1)^k$. Concretely, all the minors have a row of $-y$ across the bottom and a diagonal of $y$ with the exception of a single element where the diagonal is zero. For example, minor $M_{(n-1),(n+1)}$ has the following structure
\[M_{(n-1),(n+1)} = \begin{pmatrix}
y & 0 & 0 & \cdots & 0 & 0  \\
0 & y & 0 & \cdots & 0 & 0  \\
0 & 0 & y & \cdots & 0 & 0  \\
\vdots & \vdots & & \ddots & \vdots & \vdots \\
\vdots & \vdots & & & 0 & y  \\
-y & -y & & \cdots & -y & -y
\end{pmatrix}
\]
which is equivalent to $M_{n,(n+1)}$ with a single column swap (the last two columns). This implies $\vert M_{(n-1),(n+1)}\vert = (-1)\vert M_{n,(n+1)} \vert = y^n$. Similarly, the minor $M_{i,(n+1)}$ requires $n-i$ column swaps to move the zero column all the way to the right, which implies for $i=1,\ldots,(n-1)$ that $\vert M_{i,(n+1)}\vert = (-1)^{n-i}\vert M_{n,(n+1)}\vert = (-1)^{n-i+1}y^n$. To help see this, consider the following minor
\[M_{2,(n+1)} = \begin{pmatrix}
y & 0 & 0 & \cdots & 0 & 0  \\
0 & 0 & 0 & \cdots & 0 & 0  \\
0 & 0 & y & \cdots & 0 & 0  \\
\vdots & \vdots & & \ddots & \vdots & \vdots \\
\vdots & \vdots & & & y & 0  \\
-y & -y & & \cdots & -y & -y
\end{pmatrix}
\]
Obtaining equivalence to $M_{n,(n+1)}$ requires swapping the second column with all the $n-2$ columns to the right so $\vert M_{2,(n+1)}\vert = (-1)^{n-2+1}y^n$. Therefore, the overall expression for the jacobian simplifies to
\begin{align*}
\vert A\vert &= \left(\sum_{i=1}^n (-1)^{i+(n+1)}z_i (-1)^{n-i+1}y^n\right)+(-1)^{(n+1)+(n+1)}\left(1-\sum_{i=1}^n z_i\right)y^n\\
&= y^n\left[\left(\sum_{i=1}^n (-1)^{2n+2} z_i\right)+(-1)^{2n+2}\left(1-\sum_{i=1}^n z_i\right)\right]\\
&= y^n
\end{align*}
because $2n+2$ is an even number for any integer $n$, and $(-1)$ raised to any even power equals one.

We can now substitute this into the expression for the joint PDF of $(\vec{Z},Y)$.
\begin{align*}
f_{\vec{Z},Y}(\vec{z},y) &= f_{\vec{X}}\big(h(\vec{z},y)\big)\vert A \vert\\
&= \frac{\beta^{\alpha_0}}{\prod_{i=1}^{n+1} \Gamma(\alpha_i)} e^{-\beta y}\prod_{i=1}^n (y z_i)^{\alpha_i-1}\left(y\left(1-\sum_{i=1}^n z_i\right)\right)^{\alpha_{n+1}-1} y^n\\
&= \frac{\prod_{i=1}^n z_i^{\alpha_i-1}}{\prod_{i=1}^{n+1} \Gamma(\alpha_i)} \left(1-\sum_{i=1}^n z_i\right)^{\alpha_{n+1}-1}\beta^{\alpha_0}e^{-\beta y} y^{\sum_{i=1}^n\alpha_i-n}y^{\alpha_{n+1}-1} y^n\\
&= \frac{\prod_{i=1}^n z_i^{\alpha_i-1}}{\prod_{i=1}^{n+1} \Gamma(\alpha_i)} \left(1-\sum_{i=1}^n z_i\right)^{\alpha_{n+1}-1}\beta^{\alpha_0}e^{-\beta y} y^{\alpha_0-1}
\end{align*}
The marginal PDF of $\vec{Z}$ is obtained by integrating out $y$, producing
\begin{align*}
f_{\vec{Z}}(\vec{z}) &= \int_0^\infty f_{\vec{Z},Y}(\vec{z},y) dy = \frac{\prod_{i=1}^n z_i^{\alpha_i-1}}{\prod_{i=1}^{n+1} \Gamma(\alpha_i)} \left(1-\sum_{i=1}^n z_i\right)^{\alpha_{n+1}-1}\int_0^\infty\beta^{\alpha_0}e^{-\beta y} y^{\alpha_0-1}dy\\
&= \frac{\Gamma(\alpha_0)}{\prod_{i=1}^{n+1} \Gamma(\alpha_i)} \left(\prod_{i=1}^n z_i^{\alpha_i-1}\right)\left(1-\sum_{i=1}^n z_i\right)^{\alpha_{n+1}-1}
\end{align*}
This is immediately recognizable as the PDF of the Dirichlet distribution \citep{bela_introduction_2010},\citep{devroye_non-uniform_1986} p. 593-594, \citep{hogg_introduction_2012} p. 163-164. Although the Dirichlet has $n+1$ parameters its support is a simplex with $n$ degrees of freedom. The extra parameter can be considered a measure of dispersion in the following sense. Let $\pi_i=\alpha_i/\alpha_0$ for $i=1,\ldots,n$. Then $\E[Z_i] = \pi_i$. If $\alpha_0<1$ most of the probability mass is in the corners of the simplex (overdispersion) whereas if $\alpha_0$ is large, the density function concentrates around the mean vector $\vec{\pi}=(\pi_1,\ldots,\pi_n)$.

\section{Construction of multinomial from independent Poissons}

Let $\vec{X}=(X_1,\ldots,X_n)$ be a random vector whose elements are independent Poisson random variables such that $X_i\sim Poi(\lambda_i)$. Let $M=\sum_i X_i$ and note that $M\sim Poi(\lambda_0)$ where $\lambda_0=\sum_i \lambda_i$. Consider the density function of $\vec{X}$ conditional on $M$. By Bayes' Theorem, the conditional PMF is given by the joint PMF of $\vec{X}$ and $M$ divided by the marginal PMF of $M$.
\[p_{\vec{X}\vert M}(\vec{x}\vert m) = \frac{\prod_{i=1}^n \frac{\lambda_i^{x_i} e^{-\lambda_i}}{x_i!}}{\frac{\lambda_0^m e^{-\lambda_0}}{m!}} = \frac{m!}{\prod_i x_i!}\prod_i \left(\frac{\lambda_i}{\lambda_0}\right)^{x_i}\]
This is the multinomial PMF with total count parameter $m$ and probability parameters $\pi_i = \lambda_i/\lambda_0$ for $i=1,\ldots,n$. Note that $\sum_i \pi_i=1$ as required. Hence, a multinomial distribution is equivalent to a collection of independent Poisson distributions conditioned on their sum. This suggests that under certain conditions multinomial data may be approximated by Poisson models \citep{mcdonald_poisson_1980,baker_multinomial-poisson_1994}. Such approximations have been utilized in applications such as topic modeling \citep{gopalan_scalable_2013,taddy_distributed_2015} and genomics \citep{townes_feature_2019}.

\section{Construction of Dirichlet-multinomial from independent negative binomials}

Let $\vec{X}=(X_1,\ldots,X_n)$ be a random vector whose elements are independent negative binomial random variables such that $X_i\sim NB(\alpha_i,~\theta)$ and $\E[X_i]=\alpha_i\frac{\theta}{1-\theta}$. Let $M=\sum_i X_i$ and note that $M\sim NB(\alpha_0,~\theta)$ where $\alpha_0=\sum_i \alpha_i$. Consider the density function of $\vec{X}$ conditional on $M$. By Bayes' Theorem, the conditional PMF is given by the joint PMF of $\vec{X}$ and $M$ divided by the marginal PMF of $M$.
\[p_{\vec{X}\vert M}(\vec{x}\vert m) = \frac{\prod_{i=1}^n \frac{\Gamma(\alpha_i+x_i)}{\Gamma(\alpha_i)x_i!}\theta^{x_i}(1-\theta)^{\alpha_i}}{\frac{\Gamma(\alpha_0+m)}{\Gamma(\alpha_0)m!}\theta^{m}(1-\theta)^{\alpha_0}} = \frac{\Gamma(\alpha_0)m!}{\Gamma(\alpha_0+m)}\prod_{i=1}^n \frac{\Gamma(\alpha_i+x_i)}{\Gamma(\alpha_i)x_i!}\]
This is the Dirichlet-multinomial PMF with total count parameter $m$ and concentration parameters $\alpha_i$ for $i=1,\ldots,n$. Note that $\E[X_i\vert M] = M\frac{\alpha_i}{\alpha_0}$. Hence, a Dirichlet-multinomial distribution is equivalent to a collection of independent negative binomial distributions with the same scale parameter conditioned on their sum. This result was previously shown by \citep{zhou_nonparametric_2018}.

We note that the assumption of all the negative binomial variates having the same scale parameter ($\theta$) is crucial since otherwise the PMF of $M=\sum_i X_i$ does not have a closed form and hence neither does $\vec{X}\vert M$. This unfortunately precludes the seemingly more natural formulation of a model where $X_i$ are drawn from a negative binomial distribution with the same shape $\alpha$ but different means $\mu_i$. Such a collection of random variables, when conditioned on their sum, would not follow a Dirichlet-multinomial distribution.

\section{Discussion}

We have outlined the relationships between several distributions commonly used in modeling count data, summarized in Figure \ref{fig:diagram}. In all cases we have shown the results in closed-form. This is only possible because of conjugacy; the gamma is the conjugate prior of the Poisson and the Dirichlet is the conjugate prior of the multinomial. While conjugate priors simplify computation, they are not necessarily appropriate to all datasets. For example, the negative binomial model assumes a quadratic relationship between mean and variance. If this assumption is violated, the model will be a poor fit.

With modern computational tools, applied researchers can benefit from exploring a wider variety of compound distributions to better fit their data. For example, by replacing the gamma prior with a lognormal, one can produce a Poisson-lognormal model, which has a heavier tail than the negative binomial; this has been used for quantile normalization of single-cell gene expression data \citep{townes_quantile_2019}. Poisson-Tweedie (PT) models are another family of discrete distributions with attractive theoretical properties \citep{jorgensen_discrete_2016}. PT distributions have variance functions of the form $\mu+\phi\mu^p$, such that the negative binomial is a special case ($p=2$). While PT models can naturally handle features like zero-inflation and heavy tails, they generally do not have closed-form likelihoods which complicates their use in practical applications \citep{bonat_extended_2018}.

In the multinomial topic modeling context, a recent study found the Dirichlet prior to be overly restrictive and utilized a hierarchical nonparametric prior to improve accuracy even on data generated from a Dirichlet-multinomial model \citep{gerlach_network_2018}. Another alternative prior for the multinomial is the logistic-normal, which can handle more complex between-category correlations \citep{silverman_bayesian_2019}. In conclusion, there are fundamental theoretical connections between multinomial and count-based modeling approaches. These connections should assist practitioners in deciding whether and how to approximate computationally intractable distributions such as the Dirichlet-multinomial with simpler models such as the negative binomial.

\subsection*{Acknowledgements}

The authors thank Greg Gundersen for feedback on an early draft of the manuscript, and Barbara Engelhardt for financial support.

\bibliographystyle{plain}
\bibliography{/Users/townesf/Library/texmf/bibtex/bib/Zotero}

\end{document}